\documentclass[preprint,authoryear]{elsarticle}
\usepackage{amsmath,amssymb}
\usepackage{gensymb}
\usepackage{longtable}
\usepackage{booktabs}
\usepackage{natbib}
\usepackage{xcolor}
\usepackage{hyperref}

\usepackage{graphicx,grffile}
\makeatletter
\def\maxwidth{\ifdim\Gin@nat@width>\linewidth\linewidth\else\Gin@nat@width\fi}
\def\maxheight{\ifdim\Gin@nat@height>\textheight\textheight\else\Gin@nat@height\fi}
\makeatother
\setkeys{Gin}{width=\maxwidth,height=\maxheight,keepaspectratio}

\usepackage{lineno}
\usepackage{setspace}

\usepackage[utf8]{inputenc}

\providecommand{\tightlist}{%
  \setlength{\itemsep}{0pt}\setlength{\parskip}{0pt}}

\begin{document}

\title{Brownian-like deviation of neighboring cells in the early embryogenesis
of the zebrafish}

\author[pmc,bio]{Juan Raphael Diaz Simões}
\author[pmc]{Denis Grebenkov}
\author[bio]{Paul Bourgine}
\author[bio]{Nadine Peyriéras}

\address[pmc]{Condensed Matter Physics Laboratory, CNRS, École Polytechnique, Route de Saclay 91128 Palaiseau Cedex, France}
\address[bio]{BioEmergences Laboratory USR3695, CNRS, Université Paris-Saclay, 91198 Gif-sur-Yvette Cedex, France}

\date{\today}

\begin{abstract}
We investigate cell trajectories during zebrafish early embryogenesis
based on 3D+time photonic microscopy imaging data. To remove the
collective flow motion and focus on fluctuations, we analyze the
deviations of pairs of neighboring cells. These deviations resemble
Brownian motion and reveal different behaviors between pairs containing
daughter cells generated by cell division and other pairs of neighboring
cells. This observation justifies a common practice of using white noise
fluctuations in modeling cell movement.
\end{abstract}

\maketitle

\hypertarget{introduction}{%
\section{Introduction}\label{introduction}}

Embryogenesis is the process through which the cells of an embryo
progressively organize themselves, forming functional compartments that
carry out physiological activities \citep{fagotto}. The factors that
contribute to this development are multiple, both genetic and epigenetic
in nature \citep{holliday}. A number of models have been created aiming
to describe the dynamics of groups of cells during this process
\citep{Delile2017, Ouchi2003, Prez2007, Stott1999, VanLiedekerke2015, Zaman2005}.
Many models include a stochastic component to account for
\emph{fluctuations} over a local fluid flow. These fluctuations are
often modeled as a \emph{white noise} for its conceptual and
mathematical simplicity. White noise also allows one to for an internal
dynamics of cells, that may contribute to autonomous, undirected
displacement.

Recent progress in 3D+time imaging of embryonic development, combined to
cell-tracking algorithms \citep[\citet{Amat2014}]{faure}, made possible
the identification of cell trajectories and genealogies in developing
embryos. The zebrafish is particularly well suited for acquiring this
type of data thanks to its external development and the transparency of
its tissues. Other species that are studied using these methods are the
sea-urchin, the rabbit and the ascidian
\citep[\citet{workflow}]{Fabreges2018}. The study of these trajectories
allows us to corroborate the stochastic component of the aforementioned
models, provided that one manages to quantify trajectory fluctuations.
These fluctuations are difficult to access without a precise model of
the flow of cells during embryogenesis, which is yet not known. To
overcome this problem, we use the fact that during the early
embryogenesis of vertebrates, cells follow a flow that is continuous in
space and neighboring cells have similar trajectories. This allows us to
quantify trajectory fluctuations as the \emph{deviations} of neighboring
cells (\emph{i.e.} the difference in their positions) and their
evolution in time.

In addition, the statistics of relative positions at short times can be
used for studying the behavior of cells after division (or mitosis). It
is known that immediately after division the just divided cells are
rounder and less attached to their neighbors than other cells
\citep{cadart}. We hypothesize that this change impacts the tissue
organization. It raises also the question of the recovery of adhesion
after mitosis and whether this delay can be quantified, as pattern
reintegration by the cells after mitosis may be of importance for proper
development. We aim to quantitatively identify this difference in
attachment by analyzing the statistics of relative positions. For this
purpose, we study three groups of pairs of cells:

\begin{enumerate}
\def\labelenumi{\arabic{enumi}.}
\tightlist
\item
  neighbors: pairs of cells that are within a reference distance to each
  other (the control group);
\item
  divided neighbors: pairs composed of one cell after division and
  another cell within a reference distance of the first one;
\item
  sisters: pairs of daughter cells just after division.
\end{enumerate}

\begin{figure}
\hypertarget{fig:trajs}{%
\centering
\includegraphics[width=0.66\textwidth,height=\textheight]{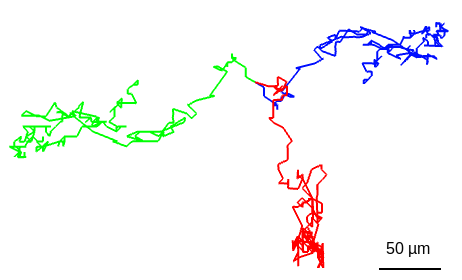}
\caption{Three sample trajectories showing the time evolution of
relative positions for neighbors (red), divided neighbors (blue) and
sisters (green) of the embryo 141108aF. The figure is a two dimensional
projection of the three dimensional trajectories. Given the statistical
variability of each of these groups, it is difficult to sample the pairs
of trajectories in a unbiased manner, which limits the representativity
of this sample.}\label{fig:trajs}
}
\end{figure}

From this definition, we see that group 1 contains group 2, that
contains group 3. The reference distance between neighbors is chosen to
match to typical distances between sisters just after division. The
study of these groups also allows us to determine possible differences
in the global deviation of cells depending on whether they are sisters
or not. In Fig.~\ref{fig:trajs} we show trajectories consisting of
relative positions of pairs of cells sampled from the three groups.

In the following we present the data, the measurements and the results.

\hypertarget{dtime-imaging-data-from-developing-embryos}{%
\section{3D+time imaging data from developing
embryos}\label{dtime-imaging-data-from-developing-embryos}}

The data consist of a spatio-temporal cell lineage algorithmically
extracted from 3D+time images of four wild type zebrafish specimens that
have been obtained by 2-photon laser scanning microscopy. Image
acquisition and image processing workflow are described in detail in
\citep{workflow}. Cell lineages have been reconstructed using the
Morphotrack cell tracking algorithm, which is a non-parametric method
based on the estimation-maximization algorithm \citep{faure}. The
characteristics of each data set are summarized in Table 1.

\begin{table}
\begin{center}
\begin{tabular}{ c c c c }
\hline
ID & Time step & Voxel size (\(\mu\)m) & Temperature (\({}^\circ\)C) \\
\hline
141108a  & 2m25 & 1.37 & 24 \\
141108aF & 2m26 & 1.38 & 28 \\
141121a  & 2m30 & 1.21 & 26 \\
170315aF & 1m40 & 1.51 & 26 \\
\hline
\end{tabular}
\caption{
Characteristics of the four studied zebrafish data sets.
The study is limited to the time interval between 6h30 and 15h15 hours post fertilization, encompassing gastrulation stages and early organogenesis.
}
\end{center}
\end{table}

\begin{figure}
\hypertarget{fig:image}{%
\centering
\includegraphics[width=1\textwidth,height=\textheight]{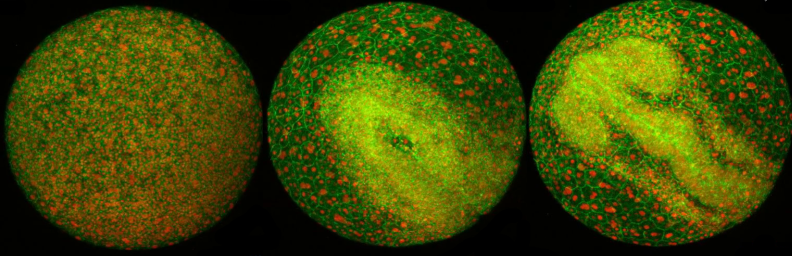}
\caption{Three snapshots of the developing zebrafish embryo
corresponding to the data set 141108aF, at 6h30 (left), 11h30 (middle)
and 16h30 (right) post fertilization respectively. Animal pole view. 3D
rendering of the raw data with nuclei stained by he expression of the
fusion between a piece of the hitsone H2B and the fluorescent protein
mCherry and membranes stained in green by the expression of a
farnesylated form of the green fluorescent protein (eGFP). On the middle
and right images one can see the result of the convergence and extension
movements, showing that cells undertake overall displacements much
larger than their size.}\label{fig:image}
}
\end{figure}

Each data set consists of 3D volumes taken at regular intervals between
6h30 and 15h15 hours post fertilization. Due to the large size of the
zebrafish embryo, the images do not encompass the animal \emph{in toto},
covering however a large portion of its body. As a consequence, cells
can enter and quit the field of view of the microscope at any moment.
Three snapshots in Fig.~\ref{fig:image} illustrate one of the data set.

The spatio-temporal lineages have been extracted from 3D images by first
detecting nucleus approximate center and linking them trough time with a
tracking algorithm \citep{workflow}, which is reported to perform with
98\% accuracy in terms of link detection between consecutive time steps.
Each data set contains between 5000 and 10000 cells per time step, the
cell number increasing in time as a consequence of cell division. The
whole dataset contains an approximate cumulated number of 100000 cells.

The reference distance for the cell neighborhood calculations has been
chosen as 20\% more than the average distance between sister cells,
corresponding to distances around 12 \(\mu\)m. Larger values for the
reference distance would lead to faster relative displacements due to
the global cell flow. Only one third of the total number of pairs of
cells for each embryo have been included into the statistics for
neighbors, to reduce the computation time. Tests using different
samplings of cells showed that this quantity of cells is enough for the
stabilization of averages, and increasing it does not change
significantly the statistics.

Given these parameters, we can count the number of cells that coexist,
\emph{i.e.} that are both present in the data set, for a period of time
larger than a certain number \(t\) of time steps, as shown for the
embryo 141108aF in Table 2. By definition, this number is decreasing,
and the decrease appears to be approximately exponential.

\begin{table}
\begin{center}
\begin{tabular}{ c c c c }
\hline
ID & Group & \# Pairs at \(t = 0\) & \# Pairs at \(t = 100\) \\
\hline
1 & neighbors & 192795 & 1725 \\
2 & divided neighbors & 156236 & 1308 \\
3 & sisters & 15395 & 122 \\
\hline
\end{tabular}
\caption{Number of pairs of cells per group that coexist for a time larger than \(t\) (for the embryo 141108aF).}
\end{center}
\end{table}

\hypertarget{measurements}{%
\section{Measurements}\label{measurements}}

We define here the measurements used in the study of deviations. In the
following \(\langle A_{ij} \rangle_{ij}\) denotes the average of the
quantity \(A\) over the set of pairs of cells denoted by \(ij\). The
error bars are calculated using the standard deviation of the quantity
\(A\) over the same set.

We denote the position of cell \(i\) at time \(t\) by
\(X_i(t) \in \mathbb{R}^3\) and define the \emph{relative position} of
two cells \(i\) and \(j\) by \[
X_{ij}(t) = X_i(t) - X_j(t)
\] which is well defined for \(t\) in the interval \([0,t_{ij}]\) at
which the two cells co-exist. In this context, time is always counted
from the first moment when both cells are present in the data set. In
other words, we do not consider the absolute time of development passed
after fertilization, cells at the beginning and end of the observation
are analyzed in the same way. Based on the relative position we define:

\begin{itemize}
\tightlist
\item
  the \emph{mean squared relative displacement} (MSRD): \[
  \delta_0(t) =
  \left\langle |X_{ij}(t) - X_{ij}(0)|^2
  \right\rangle_{ij}
  \]
\item
  the \emph{time averaged mean squared relative displacement} (TAMSRD):
  \[
  \delta_{\text{av}}(t) =
  \left\langle
  \frac{1}{t_{ij} - t + 1} \sum_{s=0}^{t_{ij} - t}|X_{ij}(s+t) - X_{ij}(s)|^2
  \right\rangle_{ij}
  \]
\end{itemize}

We also define the \emph{relative increment} \(V_{ij} \in \mathbb{R}^3\)
by: \[
V_{ij}(t) = \frac{X_{ij}(t+\Delta t) - X_{ij}(t)}{\Delta t}
\] where \(\Delta t\) is the time step (see Table 1). From it we derive
two statistics:

\begin{itemize}
\tightlist
\item
  the \emph{relative increment autocorrelation} \[
  \gamma_0(t) =
  \frac
  {\left\langle
  V_{ij}(t) \cdot V_{ij}(0)
  \right\rangle_{ij}}
  {\left\langle
  V_{ij}(0) \cdot V_{ij}(0)
  \right\rangle_{ij}}
  \]
\item
  the \emph{time averaged relative increment autocorrelation} \[
  \gamma_{\text{av}}(t) =
  \frac
  {\left\langle
  \frac{1}{t_{ij} - t + 1} \sum_{s=0}^{t_{ij} - t} V_{ij}(s+t) \cdot V_{ij}(s)
  \right\rangle_{ij}}
  {\left\langle
  \frac{1}{t_{ij} + 1} \sum_{s=0}^{t_{ij}} V_{ij}(s) \cdot V_{ij}(s)
  \right\rangle_{ij}}
  \]
\end{itemize}

As one can see from the definitions, the measurements \(\delta_0(t)\)
and \(\gamma_0(t)\) take values relative to the initial point in time
(when two cells start to coexist), while \(\delta_{\text{av}}(t)\) and
\(\gamma_{\text{av}}(t)\) include an additional time average to improve
the statistics and reduce noises. Besides better statistics, the
comparison of the two forms of measurements allows one to control the
change in dynamics over time. In particular, when a system is stationary
one should see no difference between both calculations.

\hypertarget{results}{%
\section{Results}\label{results}}

\hypertarget{short-time-dynamics}{%
\subsection{Short time dynamics}\label{short-time-dynamics}}

\begin{figure}
\hypertarget{fig:increment-all}{%
\centering
\includegraphics[width=0.66\textwidth,height=\textheight]{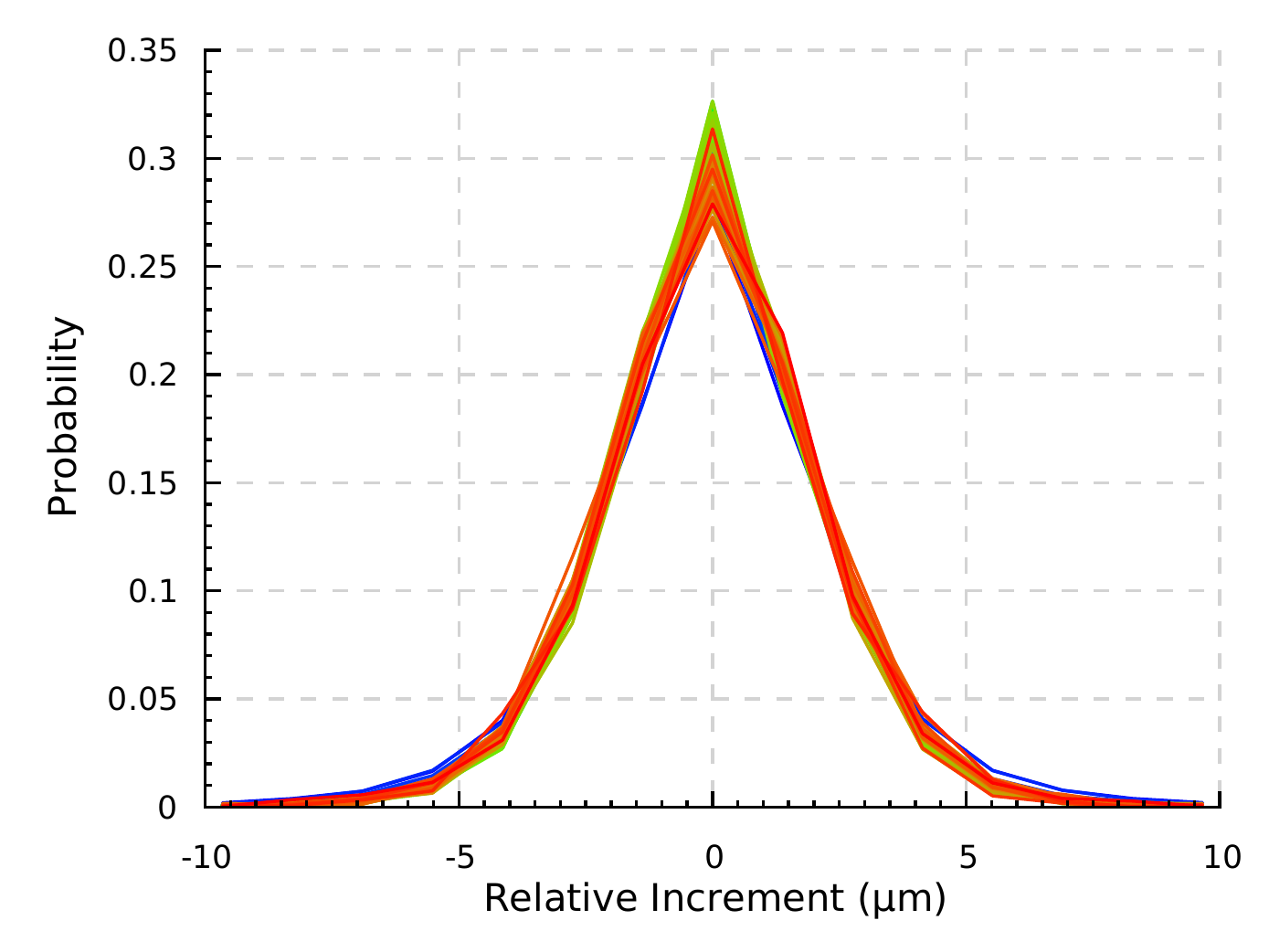}
\caption{Distributions of the \(x\) coordinate of relative increments
for pairs of group 1 (neighbors). The curves follow a color gradient
from blue to red denoting the evolution of the distribution in time, for
100 time steps (4 hours). The figure shows the relative stability of
this distribution in time. Note the discrete character of the
distribution, with a voxel size of 1.38
\(\mu\)m.}\label{fig:increment-all}
}
\end{figure}

We first study the short time dynamics by plotting the evolution in time
of the distribution of relative increments
(Fig.~\ref{fig:increment-all}). The calculation has been made for all
embryos and all three groups of cells, but since their distributions are
similar visually, we show only that of group 1 (neighbors) of the embryo
141108aF. This plot is done using the coordinate \(x\) of the
displacement vector, the result not being qualitatively different for
other coordinates. As it can be seen, the curves are fairly symmetric
and do not change much in time.

An important characteristic of this distribution is that its width is
compatible with the size of cells. That means that in general, the
relative displacement of cells between two consecutive time steps is
local and occurs in the neighborhood of the pair of cells.

Since it is difficult to evaluate the differences between the three
groups visually, we compare distributions by using the
Kolmogorov-Smirnov test \citep{corder}. This test returns a p-value for
the compatibility of two given distributions \citep{marsaglia}. For each
pair of cell groups, we applied this test at each time step. The results
were not conclusive neither for the temporal evolution of p-values nor
the relation between the three groups, for none of the embryos. In other
words, the p-values calculated both at short and long times, and for
each pair of groups were neither significantly different nor stable
enough in time to claim temporal or group differences.

\begin{figure}
\hypertarget{fig:comparison}{%
\centering
\includegraphics[width=0.66\textwidth,height=\textheight]{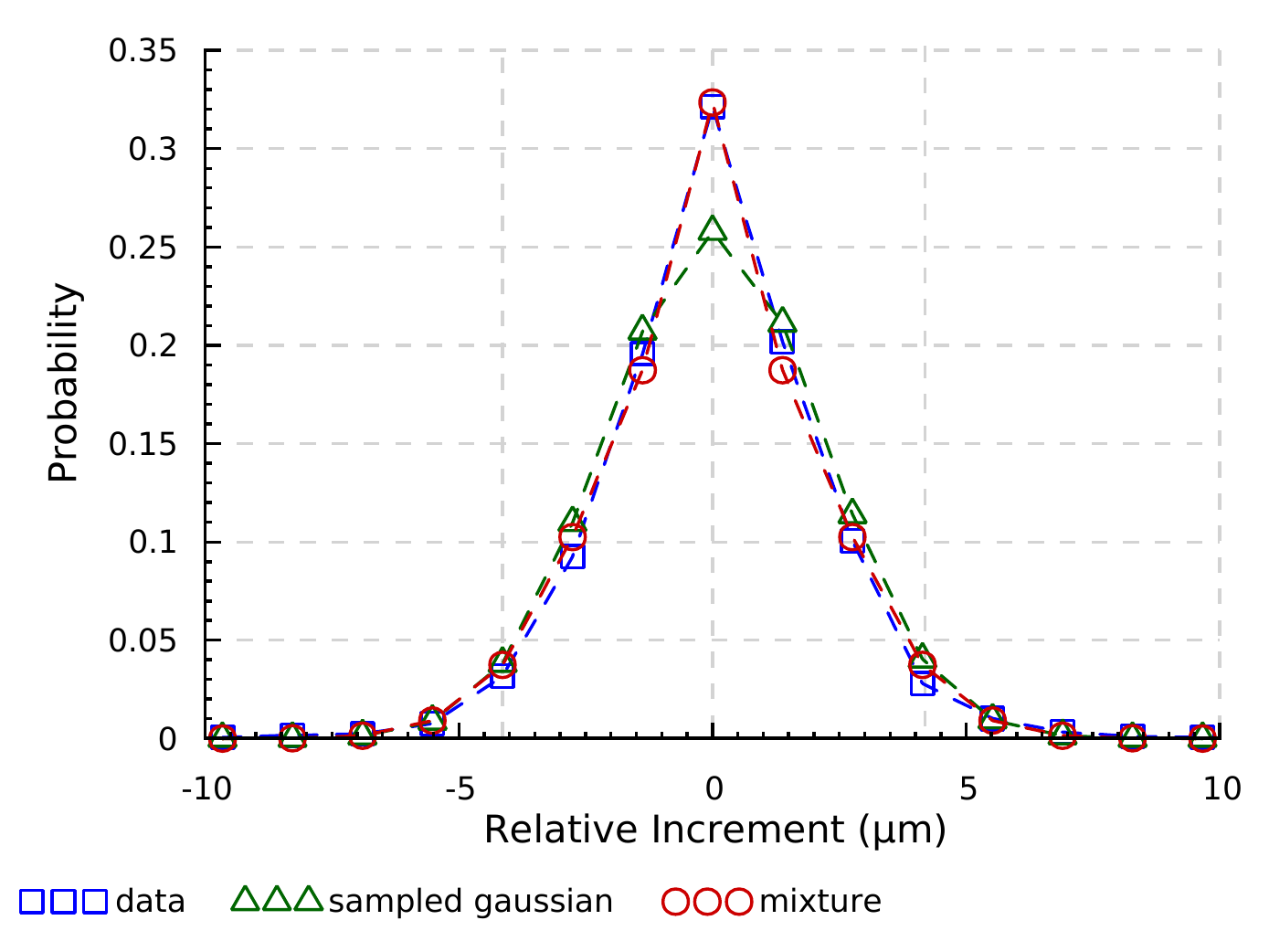}
\caption{Distributions of the \(x\) coordinate of relative increments
for pairs of neighbors at \(t = 2\) hours, with its corresponding
integer-sampled Gaussian distribution and the mixture model as defined
in Eq. (\ref{eq:mixture}) for specimen 141108aF.}\label{fig:comparison}
}
\end{figure}

We aim now to identify the distribution of relative increments
(Fig.~\ref{fig:comparison}). The first natural test is to compare this
distribution with a Gaussian one whose mean and standard deviation have
been calculated from the sample data. The Kolmogorov-Smirnov test
distinguishes the two distributions with \(p < 0.1\), which is expected
since the distribution of relative increments is discrete as a
consequence of image sampling in voxels and their relatively large size,
with 8 voxels covering all the distribution.

For this reason, we sample the Gaussian distribution accordingly. Given
a Gaussian distribution \(\mathcal{N}\), we define the integer-sampled
Gaussian distribution \(\hat{\mathcal{N}}\) by the probability
distribution:

\begin{equation}
\mathbb{P}(\hat{\mathcal{N}} = n) = \mathbb{P}\left(-\frac{1}{2} \leq \mathcal{N} - n < \frac{1}{2}\right)
\end{equation}

We compare the empirical distribution of relative increments with the
integer-sampled Gaussian one whose mean and standard deviation
parameters have been calculated from the data. Even in this case, the
Kolmogorov-Smirnov test distinguishes the two distributions for
\(p=0.1\).

As one can see in Fig.~\ref{fig:comparison}, the empirical distribution
and the integer-sampled Gaussian one differ around zero. Therefore, we
introduce the mixture distribution: \begin{equation}\label{eq:mixture}
\mathcal{M} = \alpha \hat{\mathcal{N}} + (1-\alpha)\delta_0
\end{equation} \noindent where \(\delta_0\) is a point distribution on
0. This distribution provides an effective fit that can be interpreted
as follows. There is a short-scale dynamics that cannot be accessed in
our biological data, due to the spatial granularity of the images. This
dynamics might be related to anomalous subdiffusive propagators (known
to be peaked at 0) but without having enough spatial resolution, we
model it by a Dirac delta. In addition, there is a longer-scale dynamics
which leads to a Gaussian distribution.

\begin{figure}
\hypertarget{fig:alpha}{%
\centering
\includegraphics[width=0.66\textwidth,height=\textheight]{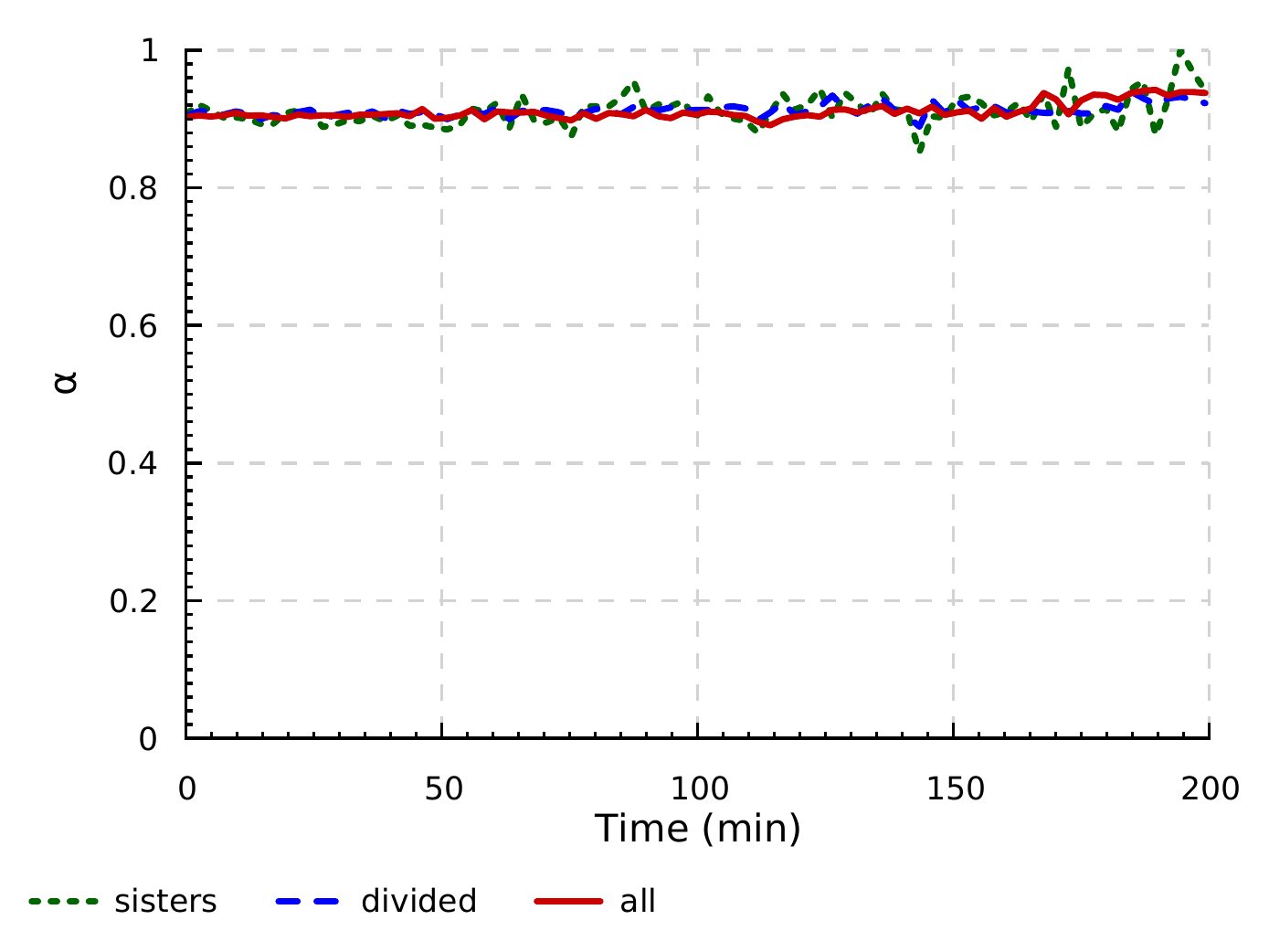}
\caption{Evolution of the parameter \(\alpha\) of the mixture model (2)
in time, for the three groups of pairs of cells.}\label{fig:alpha}
}
\end{figure}

Using a maximum likelihood estimation, we fit the distribution of
relative increments by Eq. 2. for all time steps and for all three
groups of pairs of cells. The evolution of the parameter \(\alpha\) is
shown in Fig.~\ref{fig:alpha}. As one can see, the parameter in stable
in time, fluctuating around 0.9, with no visible differences between the
three groups. Note that the mixture distribution accurately fits the
empirical one on Fig.~\ref{fig:comparison}.

\begin{figure}
\hypertarget{fig:meannorm}{%
\centering
\includegraphics[width=0.66\textwidth,height=\textheight]{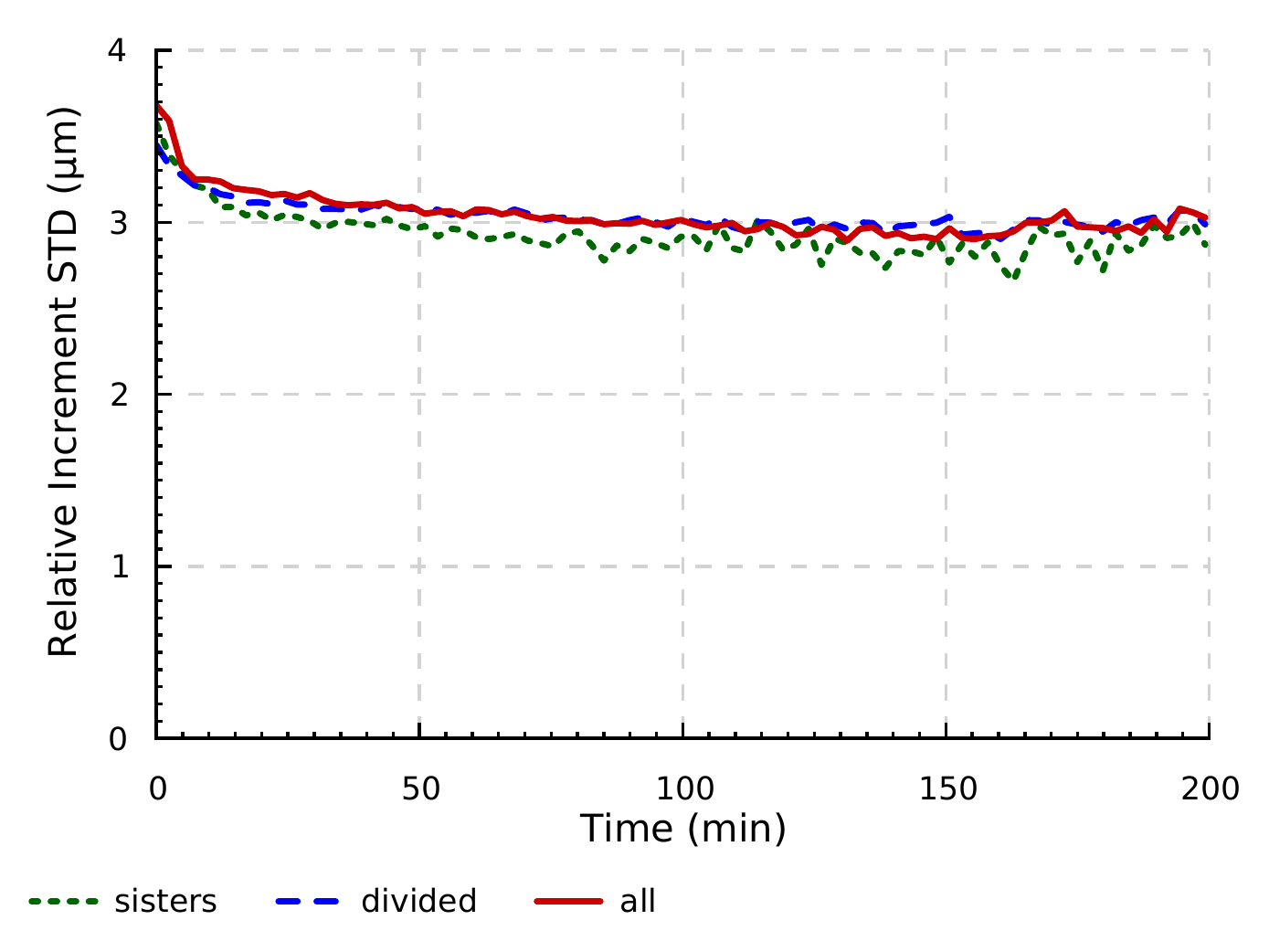}
\caption{Evolution of the standard deviation of relative increments for
the three groups: pairs of sisters, neighbors of divided cells and
neighbors for specimen 141108aF.}\label{fig:meannorm}
}
\end{figure}

The distribution of relative increments is centered around zero. We
study now the typical size of relative increments, quantified by the
standard deviation. Its time evolution is shown in
Fig.~\ref{fig:meannorm}. We can see that the standard deviation is
slightly higher at first time steps and decreases to a stable value
around 3 \(\mu m\) for all three groups. Note that this standard
deviation is smaller than the typical size of cells, which is around 10
\(\mu m\). In addition, there is no significant difference in the size
of relative increments between the three groups.

\hypertarget{long-time-dynamics}{%
\subsection{Long time dynamics}\label{long-time-dynamics}}

\begin{figure}
\hypertarget{fig:msd0}{%
\centering
\includegraphics[width=0.66\textwidth,height=\textheight]{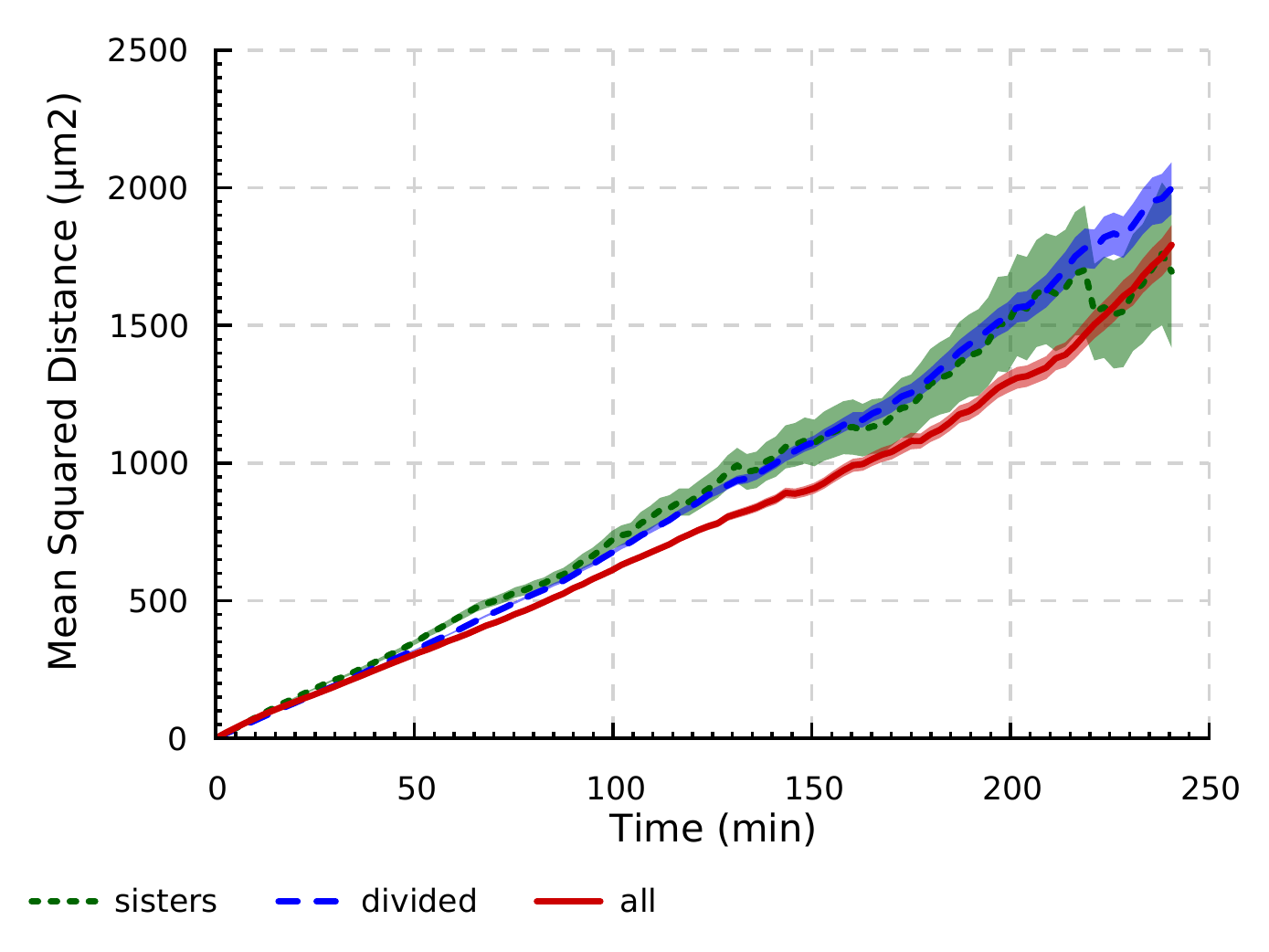}
\caption{Evolution of the mean squared relative displacement
\(\delta_0(t)\) for the three groups on specimen 141108aF. The shadowed
regions represent the standard deviation of the corresponding mean
squared relative displacement.}\label{fig:msd0}
}
\end{figure}

\begin{figure}
\hypertarget{fig:msd}{%
\centering
\includegraphics[width=0.66\textwidth,height=\textheight]{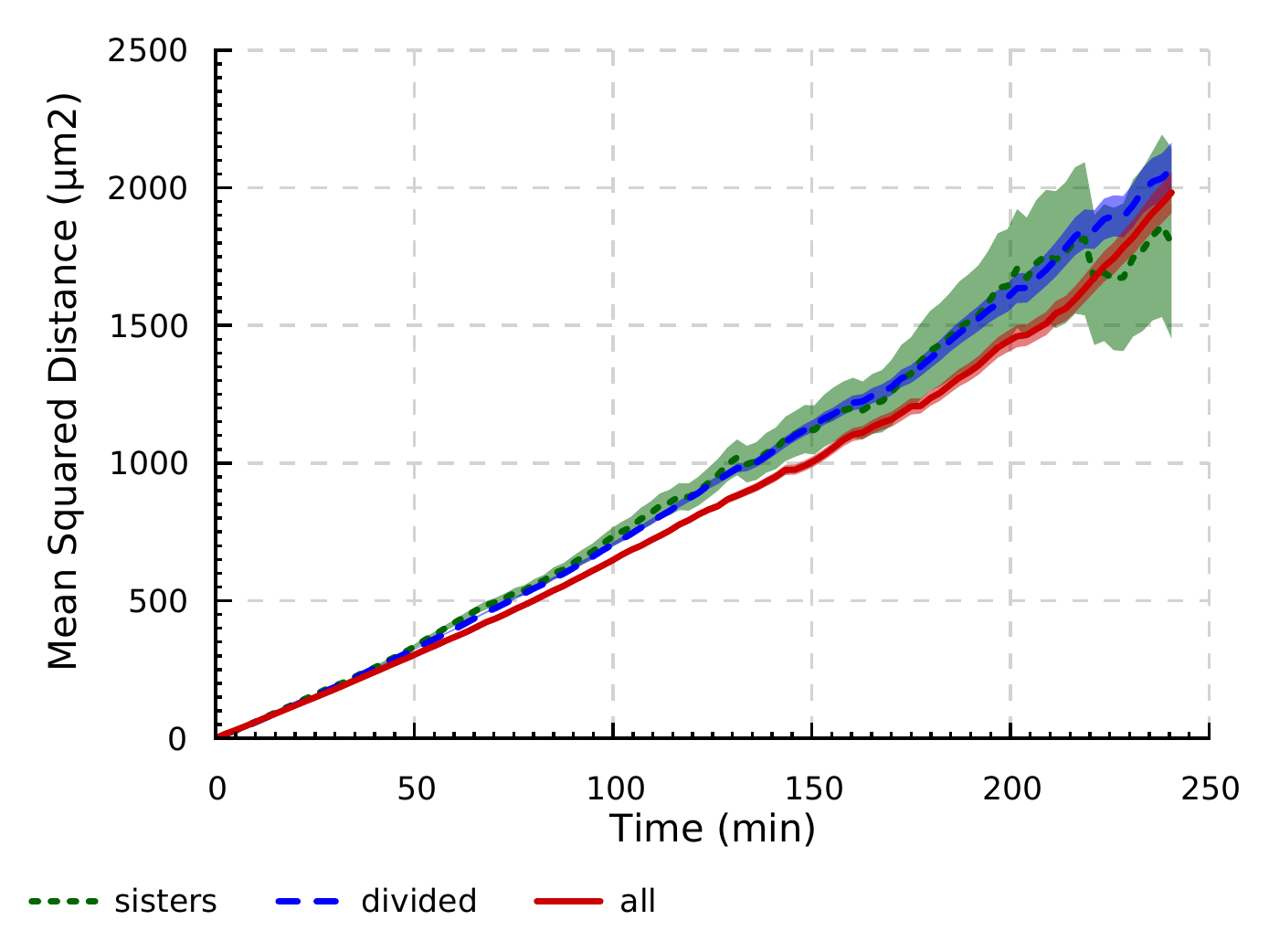}
\caption{Evolution of the time averaged mean squared relative
displacement \(\delta_{\text{av}}(t)\) for the three groups on specimen
141108aF. The shadowed regions represent the standard deviation of the
corresponding time averaged mean squared relative
displacement.}\label{fig:msd}
}
\end{figure}

We present now the evolution in time of the mean squared relative
displacements \(\delta_0\) and \(\delta_\text{av}\) for each group of
cells in Fig.~\ref{fig:msd0} and Fig.~\ref{fig:msd} for the embryo
141108aF, the result not being qualitatively different for the other
specimens.

As one can see, all these curves grow almost linearly in time. The
curves corresponding to group 1 (neighbors) have a slightly convex shape
at the end, suggesting a faster diffusion. An interpretation of this
observation is that as cells deviate from each other, effects of drift
due to differences in their respective local flows become important. It
is remarkable that the global linear behavior is observed at time scales
up to 3h, which shows a very low rate of divergence between neighboring
cells in general. The curves corresponding to the two other groups
(sisters and divided neighbors) are very similar in both cases.

\begin{table}
\begin{center}
\begin{tabular}{ c | c c | c c | c c }
\hline
Group & \multicolumn{2}{c |}{1} & \multicolumn{2}{c |}{2} & \multicolumn{2}{c}{3}\\
\hline
Measurement & MSRD & TAMSRD & MSRD & TAMSRD & MSRD & TAMSRD \\
\hline
141108a  & 0.46 & 0.47 & 0.42 & 0.46 & 0.40 & 0.46 \\
141108aF & 0.56 & 0.66 & 0.59 & 0.67 & 0.50 & 0.60 \\
141121a  & 0.45 & 0.48 & 0.40 & 0.45 & 0.37 & 0.43 \\
170315aF & 0.59 & 0.56 & 0.56 & 0.57 & 0.52 & 0.54 \\
\hline
\end{tabular}
\caption{
Diffusion coefficients (\(\mu m^2/\text{min}\)), calculated from the mean squared relative displacement \(\delta_0\) (MSRD) and time-averaged mean squared relative displacement \(\delta_\text{av}\) (TAMSRD) for each group of pairs of cells.
}
\end{center}
\end{table}

Restricting the time interval to the first 200 minutes to avoid the high
fluctuations observed in curves for group 3 (sisters), we calculate the
diffusion coefficient by fitting each curve by \(12Dt\) (the extra
factor 2 appears due to considering the relative displacement of two
cells). At 200 min we see a MSRD around 1600\(\mu m^2\), corresponding
to a relative distance of 40\(\mu m\). This shows that even if the
relative displacement between consecutive time steps is small, the
relative displacements at long times are larger than the size of the
cell, implying a change of neighborhood. An almost Gaussian character of
relative displacements and the linear growth of the MSRD with time allow
us to interpret fluctuations in the cells positions as (relative)
diffusion. We recall that this diffusion is superimposed with the
dominant coherent motion of cells.

As one can see in Fig.~\ref{fig:msd0}, the groups containing divided
cells can be distinguished from the control group at time scales up to
around 4 hours using the mean squared relative displacement and its
standard deviation. This gives a statistical significance to the
difference in diffusion coefficients (see Table 3), and supports the
observation that the relative diffusion of cells after mitosis is
slightly faster. While this observation does not depend on the specimen
being studied, the diffusion coefficient values are different by a small
margin between specimens. There are many possible reasons for this
difference, the most probable one being the difference in temperature
(Table 1), since it is known that the zebrafish embryo develops faster
at higher temperatures \citep{kimmel}, which can imply faster diffusion.
Zebrafishes develop normally between 23 and 33 degrees celsius with a
developmental speed twice higher at 33 degrees. However, we do not have
enough data to establish a causal relation between these observation.
Note also that the mean squared relative displacement \(\delta_0(t)\)
and its time averaged version \(\delta_\text{av}(t)\) yield close values
of diffusion coefficients, except for the specimen 141108aF. This embryo
was imaged at the highest temperature (see Table 1) and has the largest
deviations at later times, resulting in increased diffusion coefficients
estimated from time averages.

\begin{figure}
\hypertarget{fig:auto}{%
\centering
\includegraphics[width=0.66\textwidth,height=\textheight]{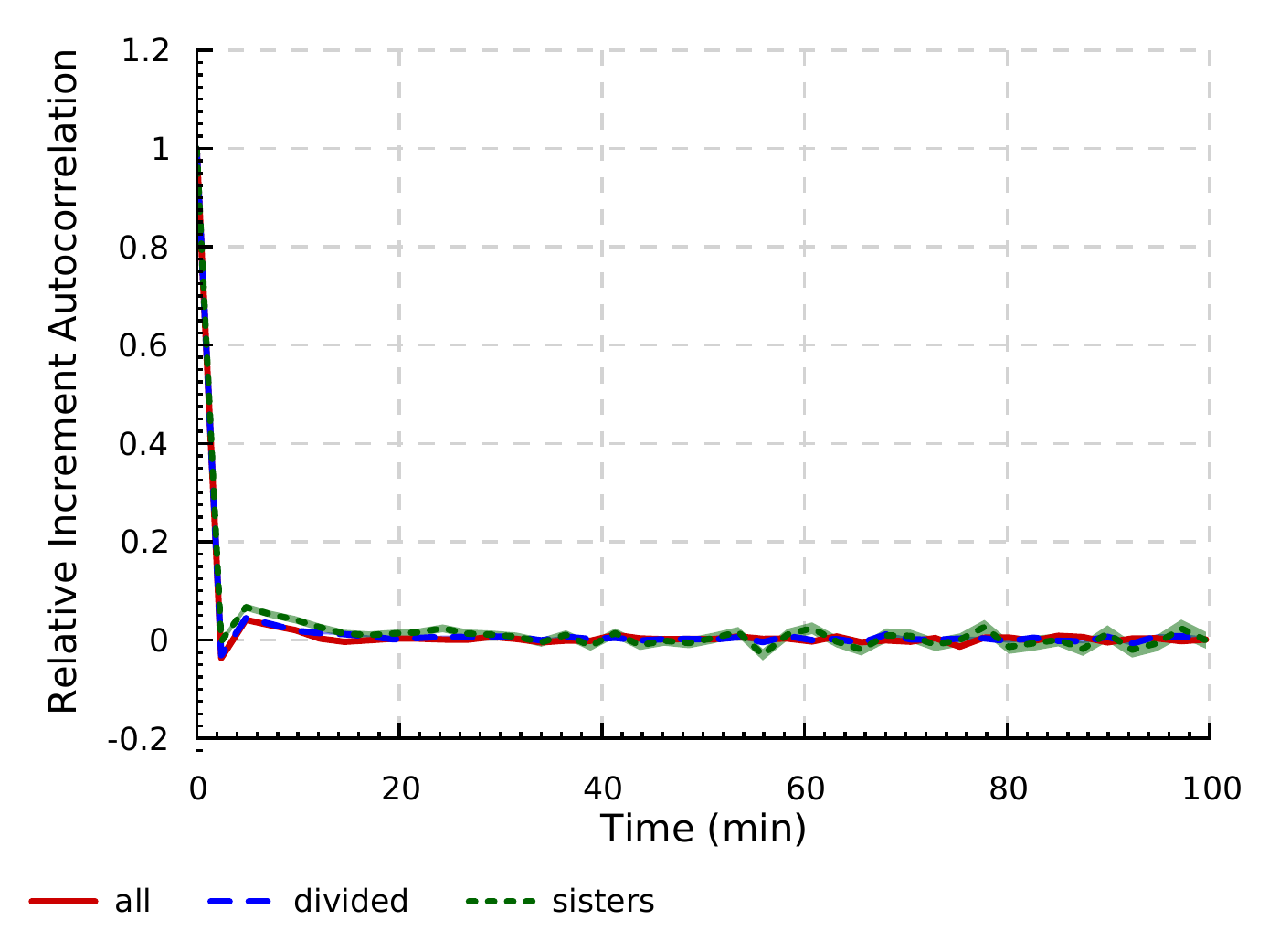}
\caption{Autocorrelation of relative increments \(\gamma_0\) for all
three groups on the specimen 141108aF.}\label{fig:auto}
}
\end{figure}

We also present the autocorrelation of relative increments \(\gamma_0\)
(Fig.~\ref{fig:auto}). For all three groups we see small negative
correlation at the first time step, and very small positive one at the
second time step. The time averaged version \(\gamma_{\text{av}}\)
behaves very similarly (not shown) and thus evidences the stationarity
of the increments of relative positions. We stress that this
stationarity does not concern the whole process of embryogenesis.

\hypertarget{conclusion}{%
\section{Conclusion}\label{conclusion}}

We reported the statistical analysis of fluctuations of cell
trajectories in the early embryogenesis of four wild type zebrafish
specimens.

The deviations between neighboring cells are very similar to Brownian
motion, and there is no visible drift effect at lag times up to 3h. This
means that, in general, neighboring cells stay close enough to each
other and remain in a region that shares a common local flow. Therefore,
one may consider the deviation between cells as a \emph{fluctuation}
around the flow, and this observation supports the stochastic component
of the models of embryogenesis dynamics
\citep{Delile2017, Ouchi2003, Prez2007, Stott1999, VanLiedekerke2015, Zaman2005}.
Furthermore, only small differences between the long-time behaviors of
the three groups of cells are observed. Therefore, the common
simplifying assumption of many models that differences in behavior of
just divided cells can be neglected is justified. The characteristics
that weakly deviate from the Brownian-like behavior are: \emph{(i)} the
relative increments are close to but not well fitted by a Gaussian
distribution (Fig.~\ref{fig:comparison}), a correction term accounting
for a short-time dynamics is proposed; \emph{(ii)} the standard
deviation of relative increments is slightly higher in the first time
steps (Fig.~\ref{fig:meannorm}); and \emph{(iii)} relative increments
are almost uncorrelated (Fig.~\ref{fig:auto}).

The calculations took into account a large proportion of the data
present in the microscope images. Therefore, the aggregated statistics
gathers cells distributed both spatially and temporally within the
explored developmental sequence. The study could be applied to selected
cells populations to explore possible spatial and temporal variations.
However, sampling the data manually and identify equivalent groups in
multiple datasets, in addition to be time consuming, may be very
inefficient. For this purpose, one could explore automated methods for
the identification of cells with similar behavior \citep{icpram}.

The data we used consisted of four different specimens of zebrafish. In
the future, it would be important to apply the same methodology to other
species in order to reveal if the relative displacement behavior
observed in our study is universal or specific to each embryogenetical
process. A potential challenge of such a study is the unequal
suitability of tracking algorithms for different species.

Finally, we have been able to identify the distribution of relative
increments as a mixture model composed of an integer-sampled Gaussian
distribution and a point distribution around zero. The former component
is related to the high discreteness of voxels, and the latter accounts
for a short time dynamics for which the current spatio-temporal
resolution of the microscopy is not sufficient. Finer details on the
distribution can potentially be obtained by using images with a higher
spatial resolution. We expect such data from recent developments in
light sheet microscopy imaging \citep{Reynaud2014}. Once combined, these
measurements can provide valuable extra information for the conception
and validation of more accurate models of embryonic development.

\hypertarget{acknowlegements}{%
\section{Acknowlegements}\label{acknowlegements}}

D.G. acknowledges the support under Grant No.~ANR-13-JSV5-0006-01 of the
French National Research Agency.

J.R.D.S. and N.P. acknowledge the support of the France BioImaging
infrastructure ANR-10-INBS-04 and the Equipex Morphoscope2
ANR-11-EQPX-0029.

\section*{References}

\bibliography{bibliography.bib}

\begin{thebibliography}{}

\bibitem[\protect\astroncite{Amat et~al.}{2014}]{Amat2014}
Amat, F., Lemon, W., Mossing, D.~P., McDole, K., Wan, Y., Branson, K., Myers,
  E.~W., and Keller, P.~J. (2014).
\newblock Fast, accurate reconstruction of cell lineages from large-scale
  fluorescence microscopy data.
\newblock {\em Nature Methods}, 11(9):951--958.

\bibitem[\protect\astroncite{Cadart et~al.}{2014}]{cadart}
Cadart, C., Zlotek-Zlotkiewicz, E., Berre, M.~L., Piel, M., and Matthews, H.~K.
  (2014).
\newblock Exploring the function of cell shape and size during mitosis.
\newblock {\em Developmental Cell}, 29(2):159--169.

\bibitem[\protect\astroncite{Corder and Foreman}{2014}]{corder}
Corder, G.~W. and Foreman, D.~I. (2014).
\newblock {\em Nonparametric Statistics: A Step-by-Step Approach}.
\newblock Wiley.

\bibitem[\protect\astroncite{Delile et~al.}{2017}]{Delile2017}
Delile, J., Herrmann, M., Peyri{\'{e}}ras, N., and Doursat, R. (2017).
\newblock A cell-based computational model of early embryogenesis coupling
  mechanical behaviour and gene regulation.
\newblock {\em Nature Communications}, 8:13929.

\bibitem[\protect\astroncite{{Diaz Sim{\~{o}}es} et~al.}{2017}]{icpram}
{Diaz Sim{\~{o}}es}, J.~R., Bourgine, P., Grebenkov, D., and Peyri{\'{e}}ras,
  N. (2017).
\newblock Cell trajectory clustering: Towards the automated identification of
  morphogenetic fields in animal embryogenesis.
\newblock In {\em Proceedings of the 6th International Conference on Pattern
  Recognition Applications and Methods, {ICPRAM} 2017, Porto, Portugal,
  February 24-26, 2017.}, pages 746--752.

\bibitem[\protect\astroncite{Fabr{\`{e}}ges et~al.}{2018}]{Fabreges2018}
Fabr{\`{e}}ges, D., Daniel, N., Duranthon, V., and Peyri{\'{e}}ras, N. (2018).
\newblock Control of the proportion of inner cells by asymmetric divisions and
  the ensuing resilience of cloned rabbit embryos.
\newblock {\em Development}, 145(8):dev152041.

\bibitem[\protect\astroncite{Fagotto}{2014}]{fagotto}
Fagotto, F. (2014).
\newblock The cellular basis of tissue separation.
\newblock {\em Development 141, 3303-3318}.

\bibitem[\protect\astroncite{Faure}{2009}]{faure}
Faure, E. (2009).
\newblock {\em {Reconstruction automatisée du lignage cellulaire
  spatio-temporel de l'embryon}}.
\newblock PhD thesis, {Ecole Polytechnique}.

\bibitem[\protect\astroncite{Faure et~al.}{2016}]{workflow}
Faure, E., Savy, T., Rizzi, B., Melani, C., Sta{\v{s}}ov{\'{a}}, O.,
  Fabr{\`{e}}ges, D., {\v{S}}pir, R., Hammons, M.,
  {\v{C}}{\'{u}}nderl{\'{\i}}k, R., Recher, G., Lombardot, B., Duloquin, L.,
  Colin, I., Koll{\'{a}}r, J., Desnoulez, S., Affaticati, P., Maury, B.,
  Boyreau, A., Nief, J.-Y., Calvat, P., Vernier, P., Frain, M., Lutfalla, G.,
  Kergosien, Y., Suret, P., Reme{\v{s}}{\'{\i}}kov{\'{a}}, M., Doursat, R.,
  Sarti, A., Mikula, K., Peyri{\'{e}}ras, N., and Bourgine, P. (2016).
\newblock A workflow to process 3d+time microscopy images of developing
  organisms and reconstruct their cell lineage.
\newblock {\em Nature Communications}, 7:8674.

\bibitem[\protect\astroncite{Holliday}{1990}]{holliday}
Holliday, R. (1990).
\newblock {DNA} methylation and epigenetic inheritance.
\newblock {\em Philosophical Transactions of the Royal Society B: Biological
  Sciences}, 326(1235):329--338.

\bibitem[\protect\astroncite{Kimmel et~al.}{1995}]{kimmel}
Kimmel, C.~B., Ballard, W.~W., Kimmel, S.~R., Ullmann, B., and Schilling, T.~F.
  (1995).
\newblock Stages of embryonic development of the zebrafish.
\newblock {\em Developmental Dynamics}, 203(3):253--310.

\bibitem[\protect\astroncite{Liedekerke et~al.}{2015}]{VanLiedekerke2015}
Liedekerke, P.~V., Palm, M.~M., Jagiella, N., and Drasdo, D. (2015).
\newblock Simulating tissue mechanics with agent-based models: concepts,
  perspectives and some novel results.
\newblock {\em Computational Particle Mechanics}, 2(4):401--444.

\bibitem[\protect\astroncite{Marsaglia et~al.}{2003}]{marsaglia}
Marsaglia, G., Tsang, W.~W., and Wang, J. (2003).
\newblock Evaluating {K}olmogorov's distribution.
\newblock {\em Journal of Statistical Software}, 8(18).

\bibitem[\protect\astroncite{Ouchi et~al.}{2003}]{Ouchi2003}
Ouchi, N.~B., Glazier, J.~A., Rieu, J.-P., Upadhyaya, A., and Sawada, Y.
  (2003).
\newblock Improving the realism of the cellular potts model in simulations of
  biological cells.
\newblock {\em Physica A: Statistical Mechanics and its Applications},
  329(3-4):451--458.

\bibitem[\protect\astroncite{P{\'{e}}rez and Prendergast}{2007}]{Prez2007}
P{\'{e}}rez, M. and Prendergast, P. (2007).
\newblock Random-walk models of cell dispersal included in mechanobiological
  simulations of tissue differentiation.
\newblock {\em Journal of Biomechanics}, 40(10):2244--2253.

\bibitem[\protect\astroncite{Reynaud et~al.}{2014}]{Reynaud2014}
Reynaud, E.~G., Peychl, J., Huisken, J., and Tomancak, P. (2014).
\newblock Guide to light-sheet microscopy for adventurous biologists.
\newblock {\em Nature Methods}, 12(1):30--34.

\bibitem[\protect\astroncite{Stott et~al.}{1999}]{Stott1999}
Stott, E., Britton, N., Glazier, J., and Zajac, M. (1999).
\newblock Stochastic simulation of benign avascular tumour growth using the
  potts model.
\newblock {\em Mathematical and Computer Modelling}, 30(5-6):183--198.

\bibitem[\protect\astroncite{Zaman et~al.}{2005}]{Zaman2005}
Zaman, M.~H., Kamm, R.~D., Matsudaira, P., and Lauffenburger, D.~A. (2005).
\newblock Computational model for cell migration in three-dimensional matrices.
\newblock {\em Biophysical Journal}, 89(2):1389--1397.

\end{thebibliography}

\end{document}